\documentclass{appolb}
\usepackage{epsfig}

\begin{document}
\title{Influence of the eta exchange to the eta production in proton-proton scattering 
}
\author{Sa\v sa Ceci, Alfreed \v Svarc, and Branimir Zauner
\address{Rudjer Boskovic Institute, Bijeni\v cka 54, HR-10000 Zagreb, Croatia}
}
\maketitle
\begin{abstract}
Eta meson production in the proton-proton scattering is dominated by the low-mass meson exchange. We present a brief study on how the type of the exchanged mesons within coupled-channel and multi-resonance model influences the scattering observables. We show under which circumstances the eta exchange may explain the shape of the observed cross sections, and present a few selected results: total cross section in the full energy range, and the proton-proton energy distribution at 15.5 MeV. 
\end{abstract}
\PACS{PACS numbers come here}
  
\section{Introduction}
Eta production in the proton-proton scattering is a well measured reaction. We have datasets of total cross sections ranging from as low as 5 MeV of excess energy, to well above 1 GeV \cite{txsdata}. The main characteristic of the data shown in Fig. 1 is its straight-line-like assembly. This arrangement is in the accordance to what would be expected from the phase space approximation: total cross section is proportional to the square of the excess energy Q. In the log-log graphical representation this is represented by the straight line. This indicates that there will be modest influence to the shape of the cross section from the different scattering amplitudes. 

\begin{figure}[!h]
\begin{center}
 \includegraphics[width=8cm]{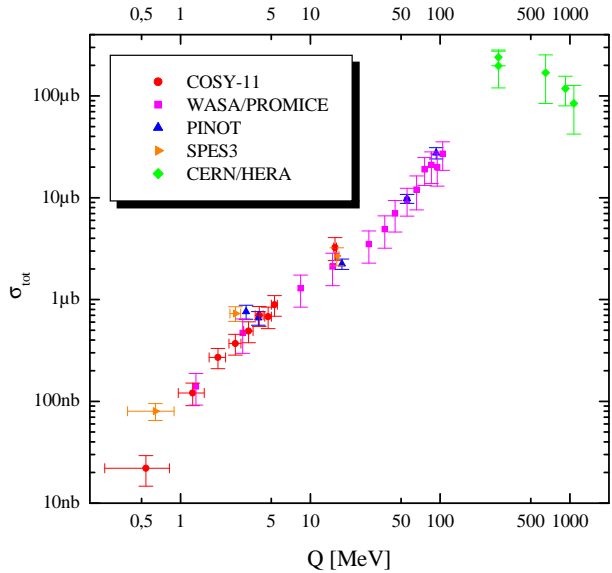}
\caption{Total cross section data for $pp\rightarrow pp\eta$. Q represents the excess energy.}\label{expdata}\end{center}
\end{figure}

There are three major contributions to the full scattering amplitude: initial state interaction (ISI) between two incoming protons, two-body off-shell eta-production amplitude (exchanged meson excites the other proton which then emits eta meson), and final state interaction (FSI) between all three particles (two protons and eta meson). 

Eta-production two-body amplitude was the one thing we readily had: it was obtained in a coupled-channel multi-resonance model \cite{Bat95} based on Cutkosky CMB approach \cite{Cut79}. The main results of our analysis were $\pi N\rightarrow\eta N$ and $\eta N\rightarrow\eta N$  scattering amplitudes.  We were motivated to do this analysis in order to test our findings about the eta-nucleon amplitudes. Given the model, obtaining the off-shell amplitude was not too much of a problem: in the relations, we used the off-shell energy and momentum of the exchanged meson. 

On the other hand, there was no simple way to include the initial and final state interaction within our model.  Therefore, in our first analysis of this process \cite{Bat98}, we have chosen the approach similar to the one used in the pion-production processes: we used only proton-proton ISI and FSI. The rationale behind the used ISI and FSI models was the change of the angular number in the proton-proton subsystem before and after the collision, since there will be a change from p-state to s-state close to threshold due to the negative parity of the eta meson. It turned out that the proton-proton ISI and FSI was enough for the description of the low-energy total cross section.

In our latest analysis \cite{Cec06}, we needed ISI and FSI that would work at higher energies as well so we could expand the energy range as much as possible. There was neither ISI nor FSI that would work at the excess energy as large as 1 GeV, at least not those that we would be able to use. There was a Jost function prescription we eventually used for proton-proton FSI [6], which asymptotically approaches to one at higher energies. Since we were interested in higher energies, we similarly assumed ISI has the value of one at high energies, but we did not model it at low energies. Since in our previous work we used ISI that was approximately a constant factor equal to 0.25 at low energies, this assumption would drastically change the low-energy behavior of our result, but at this point we were more interested in high energies. We also included the eta-proton FSI to see whether it could reduce the discrepancies between the results of our previous model and observed differential cross sections.

\section{Model}
All aspects of the model are given in detail in \cite{Cec06}. Here, we just repeat its most important characteristics.

In the full production amplitude, we considered only the exchange of pion and eta meson between two coliding protons. Exchanged meson, pion or eta, hits a proton and excites N* resonance, which decays back to proton and produces an eta meson. The crucial decision we had to make in this model was the relative sign between $\pi N \rightarrow \eta N$  and $\eta N \rightarrow \eta N$ amplitudes. The relative sign between the two amplitudes is a free parameter that cannot be determined from the partial-wave analysis.

In our previous work, we assumed constructive interference between the two amplitudes, which was enough to explain low energy size and shape of total cross section. When higher energies were considered, overall magnitude was fine. However, we realized that we cannot reproduce the shape (see \cite{Cec06}). We decided to try destructive interference, and keep ISI equal to one. Final state interaction was calculated in a multiplicative approximation: total FSI function is given by the product of Jost functions for $pp$ and $p\eta$ subsystems (as in \cite{Jost}). 

We also tried to estimate how important are resonances other than the dominant one, N(1535). Since our group is particularly interested in the research on the properties of N* resonances, we included all available resonances and all partial waves we had in our eta production amplitude.

\section{Results and discussion}
\begin{figure}[!h]\begin{center}
 \includegraphics[width=8cm]{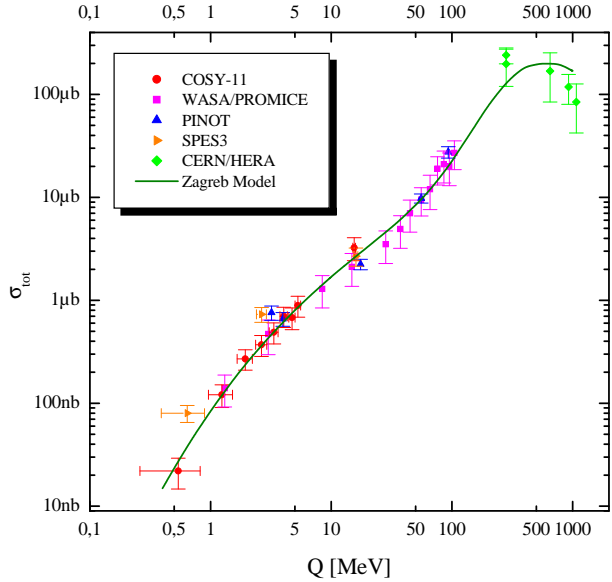}
\caption{Predicted total cross section for $pp\rightarrow pp\eta$ in a  model with all meson-nucleon partial waves included, and with the destructive interference between $\pi N$ and $\eta N$ amplitudes.}\label{results1}\end{center}
\end{figure}

Surprisingly, the change of the relative sign between the two amplitudes managed to describe the shape of the total cross section almost perfectly (Fig. \ref{results1}). The size however should still be modified by the missing ISI factor. In addition, higher partial waves of our meson-nucleon amplitude affected the results almost insignificantly below 200 MeV exces energy (c.f. \cite{Cec06}). There is a potentially interesting region with almost no data between roughly 100 and 500 MeV where higher partial-waves start to play an important role.

\begin{figure}[!h]\begin{center}
 \includegraphics[width=12cm]{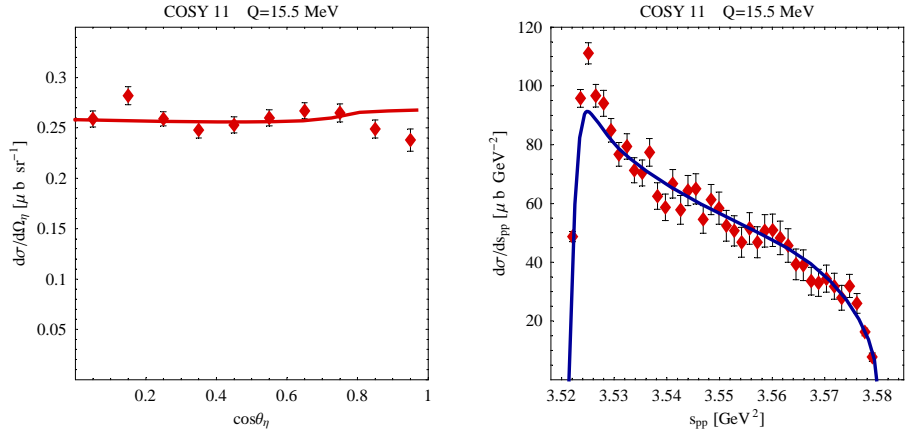}
\caption{Normalized differential cross section for $pp\rightarrow pp\eta$. Angle $\theta_\eta$ is the eta meson c.m.s. angle, while $s_{pp}$ is the square of the proton-proton invariant mass. Data are from COSY-11 \protect\cite{Pawel}.}\label{dxsPawel}\end{center}
\end{figure}

There is another somewhat unexpected result we obtained: our predictions for the shape of the differential cross sections at 15.5 MeV are in a good agreement with COSY-11 experiment \cite{Pawel}. As seen in Fig. \ref{dxsPawel}, normalized angular distribution can hardly be distinguished from the flat line, while the proton-proton energy distribution shows enough structure to be interesting to analyze.

For the analysis of the differential cross sections we used only $S_{11}$ partial wave for meson-nucleon amplitude for simplicity. The only notable difference is seen above 200 MeV for the total cross sections, and in the angular distribution for differential cross section. Namely, angular distribution still remains basically flat, just a slightly more rippled (c.f. \cite{Cec06}) than in Fig. \ref{dxsPawel}.   
 
\begin{figure}[!h]\begin{center}
 \includegraphics[width=8cm]{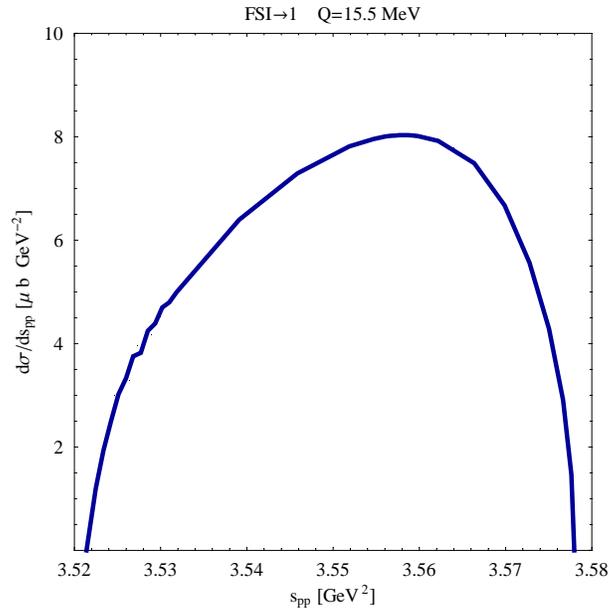}
\caption{Proton-proton energy distribution with all of the FSI contributions set equal to one. A phase space calculation would suggest symmetrical semi-elliptic shape.}\label{nofsi}\end{center}
\end{figure}

There has been lots of controversy regarding the exact proton-proton energy distribution shape, since many models have difficulties in describing it.  We tried to find out whether our shape is given by the particular choice of the  $pp$ and $\eta p$ final state, or was it in the amplitude itself. Phase space for the proton-proton energy distribution at low energies is known to be almost exactly symmetrical semi-ellipse. Proton-proton FSI produces the sharp peak at low invariant mass (close to the proton-proton threshold). We calculated this energy distribution without the final state contributions, and the result is given in Fig. \ref{nofsi}.

Evidently, our model has an asymmetrical shape of the energy distribution even without the FSI contribution. Moreover, the shape is deformed towards higher energies. Both effects, the proton-proton FSI, and eta-production amplitude are responsible for the full shape of the observed energy distribution.

\section{Conclusions}
The pi and eta meson exchange with a destructive interference were enough to reproduce the shape of the total and differential cross section. The absolute size still needs to be obtained. The low energy part of the controversial proton-proton energy distribution is dominated by the proton-proton FSI, while the eta-production amplitude pushes-up the high energy part.

Additional research is needed with the inclusion of all important contributions: namely, the rho exchange, proper ISI treatment, and detailed analysis of $pn\rightarrow pn\eta$, $pn\rightarrow d\eta$ and newly acquired  $pp\rightarrow pp\eta$ data.


\end{document}